\documentstyle{mn}
\input psfig.tex
\def\lsim{\lower.5ex\hbox{$\; \buildrel < \over \sim \;$}}
\def\gsim{\lower.5ex\hbox{$\; \buildrel > \over \sim \;$}}
\def \simeq{\lower.3ex\hbox{$\; \buildrel \sim \over - \;$}}
\def\ch{\lower-0.55ex\hbox{--}\kern-0.55em{\lower0.15ex\hbox{$h$}}}
\def\lh{\lower-0.55ex\hbox{--}\kern-0.55em{\lower0.15ex\hbox{$\lambda$}}}
\newif\ifAMStwofonts
\ifoldfss
  \ifCUPmtlplainloaded \else
    \NewTextAlphabet{textbfit} {cmbxti10} {}
    \NewTextAlphabet{textbfss} {cmssbx10} {}
    \NewMathAlphabet{mathbfit} {cmbxti10} {} 
    \NewMathAlphabet{mathbfss} {cmssbx10} {} 
  \fi
  \ifAMStwofonts
    \ifCUPmtlplainloaded \else
      \NewSymbolFont{upmath} {eurm10}
      \NewSymbolFont{AMSa} {msam10}
      \NewMathSymbol{\upi}     {0}{upmath}{19}
      \NewMathSymbol{\umu}     {0}{upmath}{16}
      \NewMathSymbol{\upartial}{0}{upmath}{40}
      \NewMathSymbol{\leqslant}{3}{AMSa}{36}
      \NewMathSymbol{\geqslant}{3}{AMSa}{3E}

    \fi
  \fi
\fi 
\ifnfssone
  \newmathalphabet{\mathit}
  \addtoversion{normal}{\mathit}{cmr}{m}{it}
  \addtoversion{bold}{\mathit}{cmr}{bx}{it}
  \newmathalphabet{\mathbfit} 
  \addtoversion{normal}{\mathbfit}{cmr}{bx}{it}
  \addtoversion{bold}{\mathbfit}{cmr}{bx}{it}
  \newmathalphabet{\mathbfss} 
  \addtoversion{normal}{\mathbfss}{cmss}{bx}{n}
  \addtoversion{bold}{\mathbfss}{cmss}{bx}{n}
  \ifAMStwofonts
    \ifCUPmtlplainloaded \else
      \UseAMStwoboldmath
      \makeatletter
      \new@mathgroup\upmath@group
      \define@mathgroup\mv@normal\upmath@group{eur}{m}{n}
      \define@mathgroup\mv@bold\upmath@group{eur}{b}{n}
      \edef\UPM{\hexnumber\upmath@group}
      \new@mathgroup\amsa@group
      \define@mathgroup\mv@normal\amsa@group{msa}{m}{n}
      \define@mathgroup\mv@bold\amsa@group{msa}{m}{n}
      \edef\AMSa{\hexnumber\amsa@group}
      \makeatother
      \mathchardef\upi="0\UPM19
      \mathchardef\umu="0\UPM16
      \mathchardef\upartial="0\UPM40
      \mathchardef\leqslant="3\AMSa36
      \mathchardef\geqslant="3\AMSa3E
    \fi
  \fi
\fi 

\ifnfsstwo
  \DeclareMathAlphabet{\mathbfit}{OT1}{cmr}{bx}{it}
  \SetMathAlphabet\mathbfit{bold}{OT1}{cmr}{bx}{it}
  \DeclareMathAlphabet{\mathbfss}{OT1}{cmss}{bx}{n}
  \SetMathAlphabet\mathbfss{bold}{OT1}{cmss}{bx}{n}
  \ifAMStwofonts
    \ifCUPmtlplainloaded \else
      \DeclareSymbolFont{UPM}{U}{eur}{m}{n}
      \SetSymbolFont{UPM}{bold}{U}{eur}{b}{n}
      \DeclareSymbolFont{AMSa}{U}{msa}{m}{n}
      \DeclareMathSymbol{\upi}{0}{UPM}{"19}
      \DeclareMathSymbol{\umu}{0}{UPM}{"16}
      \DeclareMathSymbol{\upartial}{0}{UPM}{"40}
      \DeclareMathSymbol{\leqslant}{3}{AMSa}{"36}
      \DeclareMathSymbol{\geqslant}{3}{AMSa}{"3E}
    \fi
  \fi
\fi 

\ifCUPmtlplainloaded \else
  \ifAMStwofonts \else 
    \def\upi{\pi}
    \def\umu{\mu}
    \def\upartial{\partial}
  \fi
\fi

\title{QPO frequencies
and mass-outflow rates in black hole powered galactic microquasars}
\author[Tapas K. Das, A. R. Rao and Santosh V. Vadawale]
       {Tapas K. Das $^{1,2,3}$ A. R. Rao $^4$ Santosh V. Vadawale $^4$\\
$^1$ Division of Astronomy, Department of Physics and Astronomy,
University of California at Los
Angeles, Box 951562, Los Angeles, \\
CA 90095-1562, USA\\
$^2$ Institute of Geophysics and Planetary Physics, University of California at Los
Angeles, Box 951567, Los Angeles, CA 90095, USA\\
$^3$ Racah Institute of Physics, The Hebrew University, Jerusalem 91405,
Israel\\
$^4$ Tata Institute of Fundamental Research, Homi Bhabha Road, Mumbai 400005, India\\
tapas@astro.ucla.edu, arrao@tifr.res.in, santoshv@tifr.res.in}

\date{Accepted .
      Received ;
      in original form }
\pubyear{}
\begin{document}
\onecolumn

\maketitle

\begin{abstract}
For all available pseudo-Schwarzschild potentials, we provide a 
non-self-similar model of coupled accretion-outflow
system in connection to the Quasi Periodic Oscillation (QPO) of the 
black hole powered
galactic microquasars and the emergence of barionic jets out of these
objects. 
We use the vertically
integrated 1.5 dimensional model to describe the disc structure where the
equations of motion are written on the equatorial plane of the central
accretor, assuming the flow to be in hydrostatic equilibrium in the transverse
direction.
First we 
formulate and solve the equations governing axisymmetrically rotating, advective,
multi-transonic black hole accretion which may contain the
Rankine Hugoniot Shock Waves (RHSW), and 
then we calculate the associated QPO frequencies ${\nu}_{QPO}$
in terms of relevant
accretion parameters. We then argue that the post-shock region for such
flows may serve as an efficient source of outflow generation; and we calculate,
for {\it same} set of accretion parameters used to calculate the ${\nu}_{QPO}$s,
what fraction of the accreting material, denoted by $R_{\dot m}$, is being 
blown as shock generated outflow. In this way we theoretically 
study the relation 
between ${\nu}_{QPO}$s and $R_{\dot m}$s for 
galactic microquasars and compare our
theoretically obtained result with observational data.
\end{abstract}

\begin{keywords} 
Accretion, accretion discs --- black hole physics --- hydrodynamics
--- stars ---
winds --- outflows --- x-rays: stars\\[0.25cm]
\noindent
\underline{\bf To Appear in the Monthly Notices of Royal 
Astronomical Society (MNRAS)} \\
\end{keywords} 
\section{Introduction}
It has been established in recent years that
in order to
satisfy the inner boundary conditions imposed by the
event horizon, accretion onto black holes should
exhibit transonic properties in general; 
which further
indicates that formation of shock waves
are  possible in astrophysical fluid flows onto
galactic and extra-galactic black holes. One also
expects that shock formation in black hole accretion
might be a general phenomena because shock waves
in rotating and non-rotating flows are convincingly
able to provide an important and efficient mechanism
for conversion of significant amount of the
gravitational energy (available from deep potential
wells created by these massive compact accretors) into
radiation by randomizing the directed infall motion of
the accreting fluid. Hence shocks possibly play an
important role in governing the overall dynamical and
radiative processes taking place in astrophysical fluid and 
plasma accreting 
onto black holes.
Thus the study of steady, standing, stationary shock waves produced in black
hole accretion has acquired a very important status in recent
years and it is now believed that shocks may be an
important ingredient in an accreting black hole
system in general (for an extended list of literatures on shock formation
in black hole accretion, see Das 2002, D02 hereafter, and references therein).
Hot, dense and
exo-entropic post-shock regions in advective accretion
disks are also used as a powerful tool in understanding the
spectral properties of black hole candidates (Shrader \& Titarchuk 1998, and
references therein)
and in theoretically explaining a
number of diverse phenomena, including the
generation mechanism for high frequency Quasi Periodic
Oscillations (QPOs) in
general (Titarchuk, Lapidus \& Muslimov 1998, and
references therein). A number of
observational evidences are also present which are in
close agreement with the theoretical predictions
obtained from shocked accretion model (Rutledge et al.
1999; Muno, Morgan \& Remillard 1999;
Webb \& Malkan 2000; Rao, Yadav \&
Paul 2000; Smith, Heindl \& Swank 2001).
In their attempt to explain
some of the observational features of galactic microquasars,
Chakrabarti \& Manickam (Chakrabarti \& Manickam 2000)
proposed that the intermediate frequency (${\nu}{\sim}1-10 HZ$) 
and high frequency QPO of the black hole 
candidate 
(harbored by GRS 1915+105) 
may occur due to 
oscillations of shocks in accretion disc.
In a recent work, Das (2003, hereafter D03), showed that the QPOs 
in galactic sources are, {\it indeed}, regulated by shocked accretion flow, and analytically
calculated the QPO frequency as a function of fundamental 
accretion parameters. 
\noindent
It is now a well established fact that quasars and microquasars suffer
mass loss through outflows and jets
(Mirabel \& Rodriguez 1999; Ferrari 1998; Begelman, Blandford \& Rees
1984). These galactic and extra-galactic
jet sources are commonly believed to harbor accreting compact objects
at their hearts as the prime movers for almost all non stellar energetic
activities around them including the production of bipolar outflows and
relativistic jets. Unlike normal stellar bodies, compact objects do not have
their own physical atmosphere from where matter could be ripped off as winds,
hence outflows from the vicinity of these prime movers 
have to be generated
only from the accreting material. So instead of separately
investigating the jets and accretion processes as two disjoint issues around
the dynamical center of the galactic and extra-galactic jet sources, it is
very necessary to study these two phenomena within the same framework and
any consistent theoretical model for jet production should explore the outflow
formation only from the knowledge of accretion parameters.
Also to be noted that while self-similar
models are a valuable first step, they can  never be the full answer,
and indeed any model which works equally well at all radii is fairly
unsatisfactory to prove its viability. Thus the preferred model
for jet formation must be one which is able to
select the specific region of jet formation.
Motivated by the above mentioned arguments, Das (Das 1998, D98 hereafter, Das
2001)
and Das \& Chakrabarti (Das \& Chakrabarti 1999, DC hereafter)
proposed a  non
self-similar analytical model  capable of self-consistently exploring the
hydrodynamic origin of accretion powered jets/outflows emanating
out from galactic and extra-galactic sources.
D98 and DC computed
 mass outflow rates from accretion disks around compact
objects, such as neutron stars and black holes. These computations
were done using combinations of exact transonic inflow and outflow
solutions which may form standing shock waves. Assuming that
the bulk of the outflow is from the shock generated 
effective boundary layers of these objects,
they 
found that the ratio of the outflow rate to the inflow rate varies
anywhere from a few percent to even close to a hundred percent
(i.e., close to disk evacuation case) depending on the initial parameters of the
disk, the degree of compression of matter near the centrifugal barrier,
and the polytropic index of the flow. Also
the exact location (distance measured from the
central accretor) of the jet launching zone has been successfully pointed
out in their work. \\
\noindent
Since D98, D03 and DC share a common underlying phenomena, the formation of
steady, standing Rankine Hugoniot shock waves (RHSW hereafter) in a
thin,
rotating, axisymmetric, inviscid steady accretion flow around black holes,
we believe that it is worth investigating whether theoretical
calculations allows one to correlate the frequencies of QPO of a
black hole candidate with the amount of barionic content of jets 
emanating from the vicinity of that black hole sitting at the heart 
of any galactic microquasar, and if so, whether such theoretical 
correlation may further be supported by present day observational 
evidences. Our approach in this paper is precisely this. First we 
formulate and solve the equations governing axisymmetrically rotating,
multi-transonic inviscid black hole accretion which may contain RHSW and 
then we calculate the associated QPO frequencies ${\nu}_{QPO}$
in terms of relevant
accretion parameters. We then  
argue that the post-shock region for such
flows may serve as an efficient source of outflow generation and we calculate,
for the {\it same} set of accretion parameters used to calculate the ${\nu}_{QPO}$s,
what fraction of the accreting material, denoted by $R_{\dot m}$, is being 
blown as shock generated wind. Ultimately we provide the correlation 
among ${\nu}_{QPO}$s and $R_{\dot m}$s and check the feasibility of our
theoretically obtained result against observational data. \\
\section{Inflow equations, multiplicity of sonic points and shock
formation}
\noindent
Rigorous investigation of the complete general relativistic
multi-transonic black hole accretion disc and wind
is believed to be extremely complicated.
At the same time it is
understood that as relativistic effects play an important role in the
regions close to the accreting black hole (where most of the
gravitational potential energy is released), purely Newtonian gravitational
potential (in the form ${\Phi}_{N}=-\frac{GM_{BH}}{r}$, where $M_{BH}$ 
is the mass of the compact accretor)
cannot be a realistic choice to describe
transonic black hole accretion in general. To compromise between the ease of
handling of a
Newtonian description of gravity and the realistic situations
described by complicated general relativistic calculations, a series of
`modified' Newtonian potentials have been introduced
to describe the general relativistic effects that are
most important for accretion disk structure around Schwarzschild and Kerr
black holes.
Introduction of such potentials allows one to investigate the
complicated physical processes taking place in disc accretion in a
semi-Newtonian framework by avoiding pure general relativistic calculations
so that
most of the features of spacetime around a compact object are retained and
some crucial properties of the analogous relativistic
solutions of disc structure could be reproduced with high accuracy.
Hence, those potentials might be designated as `pseudo-Kerr' or `pseudo-
Schwarzschild' potentials, depending on whether they are used to mimic the
space time around a rapidly rotating or non rotating/ slowly rotating
(Kerr parameter $a\sim0$) black
hole respectively.\\
\noindent
It is important to note that
as long as one is not
interested in astrophysical processes extremely close
(within $1-2~r_g$) to a black hole horizon, one may safely
use these  black hole potentials to study
accretion on to a Schwarzschild
black hole with the advantage that use of these
potentials would simplify calculations by allowing one
to use some basic features of flat geometry
(additivity of energy or de-coupling of various
energy components etc.) which is not possible for
calculations in a purely Schwarzschild metric.
Also, one
can study more complex many body problems such as
accretion from an ensemble of companions or overall
efficiency  of accretion onto an ensemble of black holes
in a galaxy  or  for studying numerical hydrodynamic accretion flows
around a black hole etc. as simply as can be done in a
Newtonian framework, but with far better
accuracy. However, one should be careful in using these
potentials to study shocked black hole
accretion because of the fact that none of the 
potentials discussed here
are `exact' in a sense that they are not directly
derivable from the Einstein equations.
These potentials
could only be used to obtain more
accurate correction terms over and above the pure
Newtonian results and any `radically' new results
obtained using these potentials should be cross-checked
very carefully with the exact general relativistic theory.\\
\noindent
Our calculations in this paper are based on four such pseudo
Schwarzschild potentials.
Explicit forms of those four potentials are
the following (see Das \& Sarkar 2001 and D02
references therein for detail discussion about various properties
of these potentials):
\begin{center}
$
\Phi_{1}(r)=-\frac{1}{2(r-1)}~;~
\Phi_{2}(r)=-\frac{1}{2r}\left[1-\frac{3}{2r}+12{\left(\frac{1}{2r}\right)}
^2\right]
$
\end{center}
$$
\Phi_{3}(r)=-1+{\left(1-\frac{1}{r}\right)}^{\frac{1}{2}}
~;~\Phi_{4}(r)=\frac{1}{2}ln{\left(1-\frac{1}{r}\right)}
\eqno{(1)}
$$ 
where $r$ is the radial co-ordinate scaled in units of
Schwarzschild radius. Hereafter,
 we will define the Schwarzschild radius $r_g$ as
$$
r_g=\frac{2G{M_{BH}}}{c^2}
$$
(where  $M_{BH}$  is the mass of the black hole, $G$
is universal gravitational
constant and $c$ is velocity of light in vacuum) so that the marginally bound
circular orbit $r_b$ and the last stable circular orbit $r_s$
take the values $2r_g$
and $3r_g$ respectively for a typical Schwarzschild black hole. Also,
total
mechanical energy per unit mass on $r_s$ (sometimes called
`efficiency' $e$) may be computed as $-0.057$ for this
case. We will use a simplified geometric unit throughout this paper where
radial distance $r$ is scaled in units of $r_g$, radial dynamical
velocity
$u$ and polytropic sound speed $a$ of
the flow is scaled in units of $c$ (the
velocity
of light in vacuum), mass $m$ is scaled in units of $M_{BH}$
and all other derived quantities would be scaled
accordingly. For simplicity, we will use $G=c=1$.\\
\noindent
Among the above potentials, $\Phi_1(r)$ was introduced by 
Paczy\'nski and Wiita (1980) which accurately reproduces the positions 
of $r_s$ and $r_b$. Also the Keplerian distribution of angular
momentum obtained using this potential is exactly same as
that obtained in pure
Schwarzschild geometry.
$\Phi_2(r)$ was proposed by 
Nowak and Wagoner (1991) to approximate
some of the
dominant relativistic effects of the accreting
black hole (slowly rotating or
non--rotating) via a modified Newtonian potential. It has the correct form of $r_s$
as well as it produces the best approximation for the
value of the
angular velocity $\Omega_s$
(as measured at infinity) at $r_s$ and the
radial epicyclic frequency $\kappa$ (for $r>r_s$).
$\Phi_3(r)$ and $\Phi_4(r)$ were proposed by 
Artemova, Bj\"{o}rnsson \& Novikov 1996, ABN hereafter,
to produce 
exactly the
same value of the free-fall
acceleration of a test particle at a given value of $r$ as is obtained
for a test particle at rest with respect to the Schwarzschild reference
frame ($\Phi_3$) and to produce 
the value of the free fall acceleration that is equal
to the value of the covariant component of the three dimensional free-fall
acceleration vector of a test particle that is at rest in the Schwarzschild
reference frame ($\Phi_4$) respectively.
Hereafter, we will denote any $i$th
potential as $\Phi_i(r)$ where $\left\{i=1,2,3,4\right\}$ corresponds to
$\left\{\Phi_1(r), \Phi_2(r), \Phi_3(r),\Phi_4(r)\right\}$ respectively.\\
\noindent
Following standard literature, we consider a thin, 
rotating, axisymmetric, inviscid steady flow in hydrostatic 
equilibrium in transverse direction. The assumption of hydrostatic
equilibrium is justified for a thin flow because for such a flow, the infall
time scale is expected to exceed the local sound crossing time 
scale in the direction transverse to the flow. The flow is also assumed to 
possess considerably large radial velocity which makes the flow `advective'
(see D02, and references therein).
The complete solutions of such a system 
(which may allow RHSW to form) require the dimensionless
equations for conserved specific energy ${\cal E}$ and angular 
momentum $\lambda$ of the accreting material and the mass conservation 
equations supplied by the transonic conditions at the sonic points and the 
Rankine Hugoniot conditions at the shock. 
The local half-thickness,
$h_i(r)$ of the disc for any $\Phi_i$ can be obtained by balancing the
gravitational force by pressure gradient and can be expressed as:
$$
h_i(r)=a\sqrt{{r}/\left({\gamma}{\Phi_i^{\prime}}\right)}
\eqno{(2)},
$$
$\Phi_i^{\prime}$ is the derivative of any
$i$th potential with respect to the radial co-ordinate.
For a non-viscous flow obeying the polytropic equation of state
$p=K{\rho}^{\gamma}$ ($K$ is a measure of
the specific entropy of the flow),  integration of radial momentum
equation:
$$
u\frac{{d{u}}}{{d{r}}}+\frac{1}{\rho}
\frac{{d}p}{{d}r}+\frac{d}{dr}\left[\Phi^{eff}_{i}(r)\right]=0
$$
where $\Phi_i^{eff}(r)$ is the $i$th `effective' potential which is 
the summation of the gravitational
potential and the centrifugal potential for matter
accreting under the influence of $i$th pseudo
potential:
$$
\Phi_i^{eff}(r)=\Phi_i(r)+\frac{\lambda^2}{2r^2},
$$
leads to the following energy conservation equation in steady state:
$$
{\cal E}=\frac{1}{2}u_e^2+\frac{a_e^2}{\gamma - 1}
+\frac{{\lambda}^2}{2r^2}+\Phi_i(r)=0;
\eqno{(3a)}
$$
Similarly, the continuity equation:
$$
\frac{{d}}{{d}r}
\left[u{\rho}rh_i(r)\right]=0
$$
can be integrated to obtain the barion number conservation equation as:
$$
{\dot M}=\sqrt{\frac{1}{\gamma}}u_ea_e{\rho}_er^{\frac{3}{2}}
\left({\Phi_i^{\prime}}\right)^{-\frac{1}{2}}.
\eqno{(3b)}
$$
Following standard literature, one can define the entropy accretion rate
${\dot {\cal M}}={\dot M}K^{\left(\frac{1}{\gamma-1}\right)}
{\gamma}^{\left(\frac{1}{\gamma-1}\right)}$
which undergoes a  discontinuous transition
at the shock location $r_{sh}$ where local turbulence generates entropy to
increase ${\dot {\cal M}}$ for post-shock flows. For our purpose,
explicit expression for ${\dot {\cal M}}$ can be obtained as:
$$
{\dot {\cal M}}=\sqrt{\frac{1}{\gamma}}u_e
a_e^{\left({\frac{\gamma+1}{\gamma-1}}\right)}
r^{\frac{3}{2}}\left({\Phi_i^{\prime}}\right)^{-\frac{1}{2}}.
\eqno{(3c)}
$$
In Eqs. (3a-3c), the subscript $e$ indicates the values measured on the
equatorial plane of the disk;
however,
we will drop $e$ hereafter if no confusion arises in doing so.\\
\noindent
At this point, we would like to discuss in detail about some of the assumptions made
in this work to study the disc structure and dynamics. Firstly, our calculations are
essentially based on inviscid accretion whereas in reality one may expect that
viscosity is present in
accretion disc around black holes. The exact expression of $\lambda$
in eq. (8a) may, thus be $\lambda(r)$ (which means that the angular momentum would be a
function of the radial distance) and along with the radial momentum equation and
continuity equation, two other equations, namely, equations for azimuthal angular
momentum and the equation for heating and cooling factors describing the radiative
properties of the flow, should be used to bring the
whole picture into focus. However, we try to justify our assumption of inviscid flow
in the following way:\\
We will mainly concentrate in a region of accretion disc close to the central
accretor (a few Schwarzschild radii  away from the event horizon). This is because,
except for a few unusual situations, for all reasonable initial boundary conditions for
flow in every $\Phi_i$, shocks are likely to form in this length scale (see DO2).
Close to the black hole at that length scale, the radial velocity of matter becomes
enormously large due to extremely strong gravitational attraction of the black hole,
hence infall time scale becomes much smaller compared to the viscous time scale for
all practical purposes. So, one can treat the angular momentum to be
practically constant irrespective of the nature of viscosity in accretion disc and
our assumption of inviscid flow may not be quite unjustified. Nevertheless, one
limitation of our model is to treat the flow to be inviscid even at far field
position, i.e, at a long distance away from the accretor where the viscous
dissipation of angular momentum may be taken into account. We thus, were
unable to prescribe any analytical means using which we could smoothly join the far
field high viscous flow with near field weakly viscous / practically inviscid flow.
However, even thirty years after the discovery of standard accretion disc theory
(Shakura \& Sunyaev 1973), exact modeling of viscous multi-transonic black hole
accretion, including proper heating and cooling mechanism is quite an arduous
task, and  we do  not attempt it in this paper. However, our qualitative
calculations show that the introduction of viscosity via a radius dependent power
law distribution for angular momentum only weakness the strength of the centrifugal
barrier and pushes the shock location closer to the event horizon, keeping
the overall basic physics concerning the shock dynamics and related
issues unaltered; details of
this work is in progress and will be discussed elsewhere.\\
\noindent
One can simultaneously solve Eqs. (3a-3c) for any particular $\Phi_i$ and for a
particular set of values of $\left\{{\cal E}, \lambda, \gamma\right\}$.
Hereafter we will use the notation $\left[{\cal P}_i\right]$ for a set of
values of $\left\{{\cal E}, \lambda, \gamma\right\}$ for any particular
$\Phi_i$.\\
\noindent
For a particular $\left[{\cal P}_i\right]$, it is now quite 
straight-forward to derive the space gradient of dynamical flow velocity
$\left(\frac{du}{dr}\right)_i$ for flow in any particular 
$i$th black hole potential $\Phi_i(r)$ as:
$$
\left(\frac{du}{dr}\right)_i=
\frac{
\left(\frac{\lambda^2}{r^3}+\Phi^{'}_i(r)\right)-
\frac{a^2}{\gamma+1}\left(\frac{3}{r}+
\frac{\Phi^{''}_i(r)}{\Phi^{'}_i(r)}\right)
}
{u-\frac{2a^2}{u\left(\gamma+1\right)}
}
\eqno{(4a)}
$$
Since the flow is assumed to be smooth everywhere, if
the denominator of eqn. (4a)  vanishes at any radial distance
$r$, the numerator must also vanish there to maintain the
continuity of the flow. One therefore arrives at the so
called `sonic point (alternately, the `critical point')
conditions'  by simultaneously making
the numerator and denominator of eqn. (4a) equal to zero.
The sonic point conditions then can be expressed as:
$$
a^i_s=\sqrt{\frac{1+\gamma}{2}}u^i_s=
\sqrt{
\left[
\frac{\Phi^{'}_i(r)+{\gamma}\Phi^{'}_i(r)}{r^2}
\left(
\frac{\lambda^2+r^3\Phi^{'}_i(r)}{3\Phi^{'}_i(r)+r\Phi^{''}_i(r)}
\right)
\right]_s
}
\eqno{(4b)}
$$
where the subscript $s$ indicates that the quantities are to be measured
at the sonic point(s) and ${\Phi_i}^{{\prime}{\prime}}$ represents 
the derivative of ${\Phi_i}^{{\prime}}$. 
For a fixed $\left[{\cal P}_i\right]$
and $\Phi_i$, one can solve the following polynomial of $r$
to obtain the sonic point(s) of the flow
where the subscript $s$ refers to the quantities measured at the
sonic points:
$$
{\cal E}-{\left(\frac{\lambda^2}{2r^2}+\Phi_i
\right)}_{s}-\frac{2\gamma}{\gamma^2-1}
\left[
\frac{\Phi^{'}_i(r)+{\gamma}\Phi^{'}_i(r)}{r^2}
\left(
\frac{\lambda^2+r^3\Phi^{'}_i(r)}{3\Phi^{'}_i(r)+r\Phi^{''}_i(r)}
\right)
\right]_s
=0.
\eqno{(4c)}
$$
Similarly, the value of $\left(\frac{du}{dr}\right)_i$ at its 
corresponding sonic point(s) $r^i_s$ can be obtained by solving the 
following equation:
$$
\frac{4{\gamma}}{\gamma+1}\left(\frac{du}{dr}\right)^2_{s,i}
-2u_s\left(\frac{\gamma-1}{\gamma+1}\right)
\left(\frac{3}{r}+\frac{\Phi^{''}_i(r)}{\Phi^{'}_i(r)}\right)_s
\left(\frac{du}{dr}\right)_{s,i}
$$
$$
+a^2_s\left[\frac{\Phi^{'''}_i(r)}{\Phi^{'}_i(r)}
-\frac{2\gamma}{\left(1+{\gamma}\right)^2}
\left(\frac{\Phi^{''}_i(r)}{\Phi^{'}_i(r)}\right)^2
+\frac{6\left(\gamma-1\right)}{\gamma{\left(\gamma+1\right)^2}}
\left(\frac{\Phi^{''}_i(r)}{\Phi^{'}_i(r)}\right)
-\frac{6\left(2\gamma-1\right)}{\gamma^2{\left(\gamma+1\right)^2}}
\right]_s
$$
$$
+
\Phi^{''}_i{\Bigg{\vert}}_s-
\frac{3\lambda^2}{r^4_s}=0
\eqno{(4d)}
$$
Where the subscript $``s,i"$ indicates that the corresponding 
quantities for any $i$th potential is being measured at its 
corresponding sonic point(s) and $\Phi^{'''}_i(r)=\frac{d^3\Phi_i(r)}{dr^3}$.\\
\noindent
\begin{figure}
\vbox{
\vskip -3.5cm
\centerline{
\psfig{file=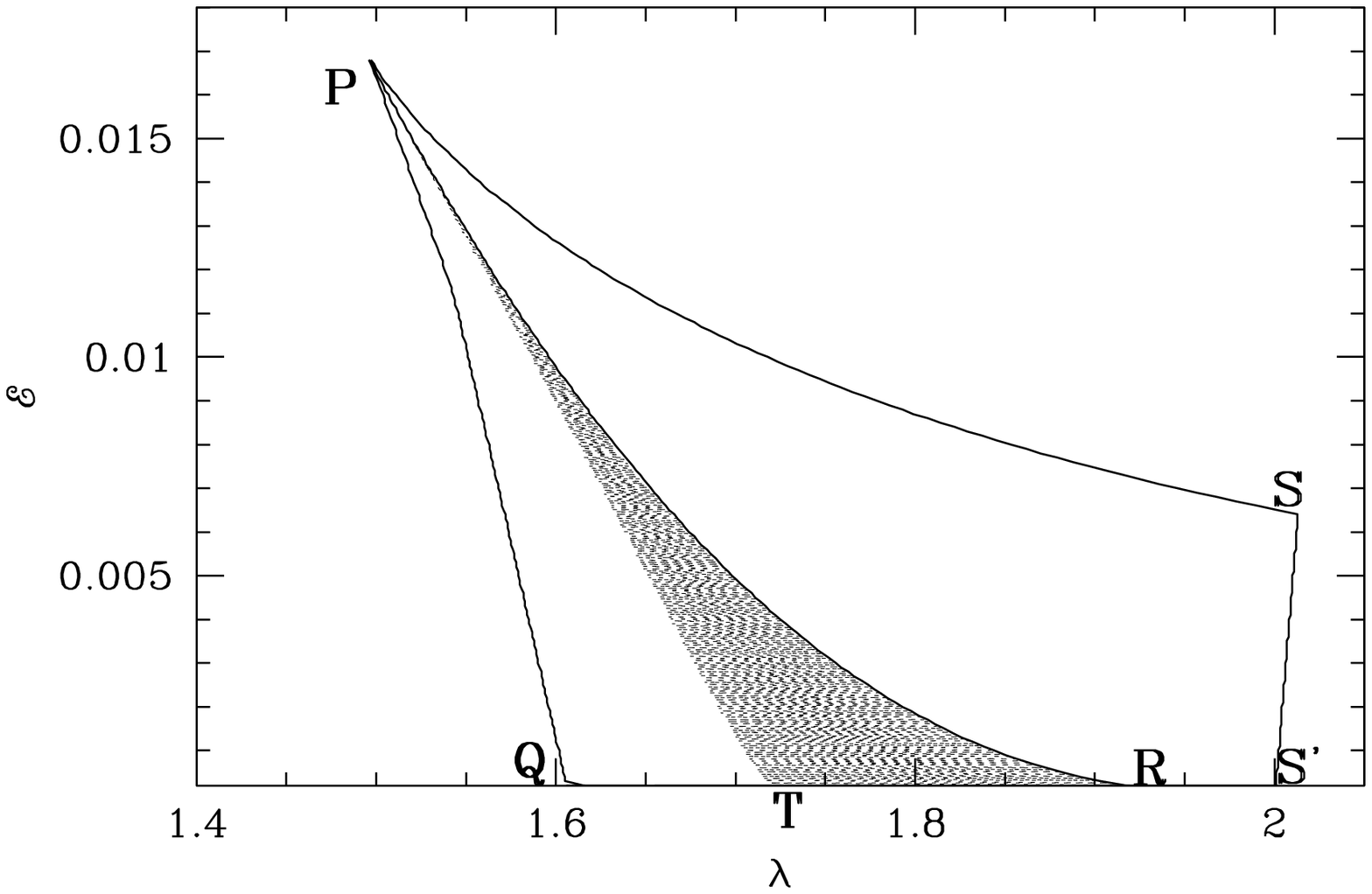,height=12cm,width=12cm,angle=0.0}}}
\noindent {{\bf Fig. 1:}
Classification of parameter space for multi-transonic accretion and wind.
Shock formation region is shown for accretion only and not for wind.
See text for details.}
\end{figure}
For {\it all} $\Phi_i$'s,
we find a significant region of
parameter space spanned by $\left[{\cal P}_i\right]$ which  allows
the multiplicity of
sonic points for accretion as well as for wind
where two real physical inner and outer (with respect to
the black hole location) $X$ type sonic points $r_{in}$ and $r_{out}$ encompass
one $O$ type unphysical middle sonic point $r_{mid}$ in between them.
If shock forms in accretion (in this work we will not study the
shock formation in wind), then $\left[{\cal P}_i\right]$s responsible
for shock formation must be somewhere from the region for which three
sonic points will form in {\it accretion only}, though 
not all $\left[{\cal P}_i\right]$s in the region of
multi-transonic accretion
will allow shock transition (see subsequent discussions and Fig. 1).
Using Eqs. (2,3a-3c), one can
combine the three standard RH conditions (Landau \& Lifshitz 1959)
for vertically integrated pressure and density
to
derive the following relation which is valid {\it only} at the shock
location: 
$$
\left(1-\gamma\right)\left(\frac{{\rho_{-}}{{\dot {\cal M}}_{-}}}{\dot M}
\right)^{log_{\Gamma}^{1-\Theta}}
{\cal E}_{{\left(ki+th\right)}}
-{\Theta}{\left(1+\Theta-R_{comp}\right)}^{-1}
+\left(1+\Theta\right)^{-1}
=0,
\eqno{(5)}
$$
where ${\cal E}_{{\left(ki+th\right)}}$ is the total specific thermal plus
mechanical energy of the accreting fluid:
${\cal E}_{{\left(ki+th\right)}}=\left[{\cal E}-
\left(\frac{\lambda^2}{2r^2}+\Phi_i\right)\right]$, $R_{comp}$ and $\beta$ are
the density compression and entropy enhancement ratio respectively, defined
as
$R_{comp}=\left({\rho_{+}}/{\rho_{-}}\right)$ and
$\beta=\left({\dot {\cal M}}_{+}/{\dot {\cal M}}_{-}\right)$
respectively; $\Theta=1-\Gamma^{\left(1-{\gamma}\right)}$ and $\Gamma={\beta}
{R_{comp}}$, ``$+$'' and ``$\_$'' refer to the post- and
pre-shock quantities.
The shock
strength ${\cal S}_i$ (ratio of the pre- to post-shock Mach number of the
flow) can be calculated as:
$$
{\cal S}_i=R_{comp}\left(1+\Theta\right).
\eqno{(6)}
$$
One can simultaneously solve Eqs. (3a-3c,5-6) to find out the
shock location 
$r_{sh}$
along with
any sonic or shock quantity as a function of
$\left[{\cal P}_i\right]$ and can identify the regions of parameter 
space (spanned by ${\cal E}$, $\lambda$ and $\gamma$) responsible 
for shock formation for any $\Phi_i(r)$.
However, one can show that as far as the shock formation in multi-transonic
accretion flows around black holes are concerned, the Paczy\'nski and Wiita
potential is expected to be the best approximation of complete general 
relativistic solutions (D02), and hence, although we will provide
a general solution scheme for shock induced outflow for {\it all}
pseudo potentials,
we will present our main results for accretion in ${\Phi_1}(r)$ only.\\
\noindent
In Fig. 1, we classify the parameter space for multi-transonic accretion 
and wind for flows in ${\Phi_1}(r)$. 
The specific
energy ${\cal E}$ is plotted along the $Y$ axis and the specific
angular momentum
$\lambda$ is plotted along the $X$ axis.
In the region bounded by {\bf PQR},
three sonic points are formed in {\it accretion},
while three sonic points in 
{\it wind} are formed in the region bounded by {\bf PRS$^{'}$S}. The shaded 
wedge {\bf PTR} represents the region of parameter space for which 
multi-transonic accretion flow will have a steady, standing RHSW. 
Any region which falls outside {\bf PQS$^{'}$S}, will produce a mono-transonic
accretion, so to say. Although the figure is drawn for 
ultra-relativistic
\footnote{By the term `ultra-relativistic' and `purely non-relativistic'
we mean a flow with
$\gamma=\frac{4}{3}$ and $\gamma=\frac{5}{3}$ respectively,
according to the terminology used in
Frank et. al (1992).} flow, one can easily explore multi-transonic 
shocked accretion flow with higher values
of $\gamma$ ($4/3<{\gamma}<5/3$).
For any $\left[{\cal P}_i\right]\in {\bf PTR}$, one can calculate the 
shock location $r_{sh}$ and the shock compression ratio $R_{comp}$.
Following D03, one can calculate the QPO frequency ${\nu}_{QPO}$
as:
$$
{\nu}_{QPO}=\frac{{\cal A}}{R_{comp}r^{\frac{3}{2}}_{sh}}
\eqno{(7)}
$$
The dimensionless constant ${\cal A}$ in the above equation can
be determined by scaling the calculated QPO frequencies with the highest 
possible {\it observed} ${\nu}_{QPO}$ for a particular astrophysical
source.
\section{Generation of shock-induced outflow and its governing equations}
\noindent
The simplicity of black holes lie in the fact that they do
not have atmospheres. But the advective accretion 
discs surrounding them have, and similar method
as employed in stellar atmospheres should be applicable to the disks.
Our approach in this section is precisely this.
We first explain how the post shock region is expected to produce outflows;
in doing so we will follow more or less the same arguments used by D98 and DC.
Then we will formulate the equations governing such outflows in a general format
so that we will incorporate all pseudo potentials. Finally we will 
simultaneously solve the equations governing the inflow and outflow to find
out what fraction of the accreting material, denoted by $R_{\dot m}$,
is being blown as wind and for the {\it same} set of accretion parameters, we
will correlate that fractional amount with the frequencies of the
QPO.
\noindent
Due to the fact that close to the black hole 
the radial component of the
infall velocity of accreting material would be enormously high,
viscous time scale would be much longer than the infall time scale and
a rotating inflow
entering into a black hole  will have almost constant
specific angular momentum
close to the black hole for any moderate viscous stress.
This almost constant angular momentum
produces a very strong centrifugal force  which
increases much faster compared to the gravitational force and becomes comparable
 at
some specific radial distance.
Here, (actually, a little farther out, due to
thermal pressure) matter starts
piling up and produces the centrifugal pressure supported boundary layer
(CENBOL). Further close to the black hole, the gravity always wins
and matter enters the horizon supersonically after passing
through a sonic point.
Formation of CENBOL may be attributed to the shock formation in accreting fluid.
In CENBOL region
the post-shock flow becomes hotter and denser
and for all practical purposes
behaves as the stellar atmosphere so far as the formation of
outflows are concerned.
A part of the hot and dense shock-compressed inflowing material
is then `squirt' as outflow from the CENBOL.
Subsonic outflows originating
from CENBOL would pass through outflow sonic points and reach far distances
as in wind solution.\\
\noindent
It is to be noted here that the generation of outflow from CENBOL in this work is a
rational assumption. The exact analytical calculation describing the change
of linear momentum of the accreting material in a direction perpendicular to the
plane of the disc is beyond the frame work of the 1.5 dimensional disc model used in
this work where the explicit variation of dynamical variables along the Z axis is not
amenable to analytical treatment.
It can be stated that the enormous post shock thermal pressure in CENBOL is, in
reality, capable of providing a substantial amount of `hard push' to the accreting
material against the gravitational attraction of the black hole and this `thermal
kick' may play an important role in re-distributing the linear momentum of the
inflow and generates a non zero component along the Z direction. In other words,
thermal pressure at CENBOL, being anisotropic in nature, may deflect a part of inflow
along the perpendicular to the equatorial plane of the disc, exact mechanism of which we could
not formulate analytically. Thus in our model, the CENBOL successfuly connects the
two stationary quasi-one dimensional solutions, namely, the 1.5 dimensions vertically
averaged inflow and the outflow, to produce a coupled disc-outflow system.\\
\noindent
However, we do believe that
the CENBOL region is generic (unless
viscosity is exorbitantly high which removes angular
momentum of the flow almost completely), because all
the self-consistent solutions of the governing equations show that
angular momentum is almost constant close to the black hole even
though the matter start with a large angular momentum at a large
(few million) Schwarzschild radius away. 
\begin{figure}
\vbox{
\vskip -0.0cm
\centerline{
\psfig{figure=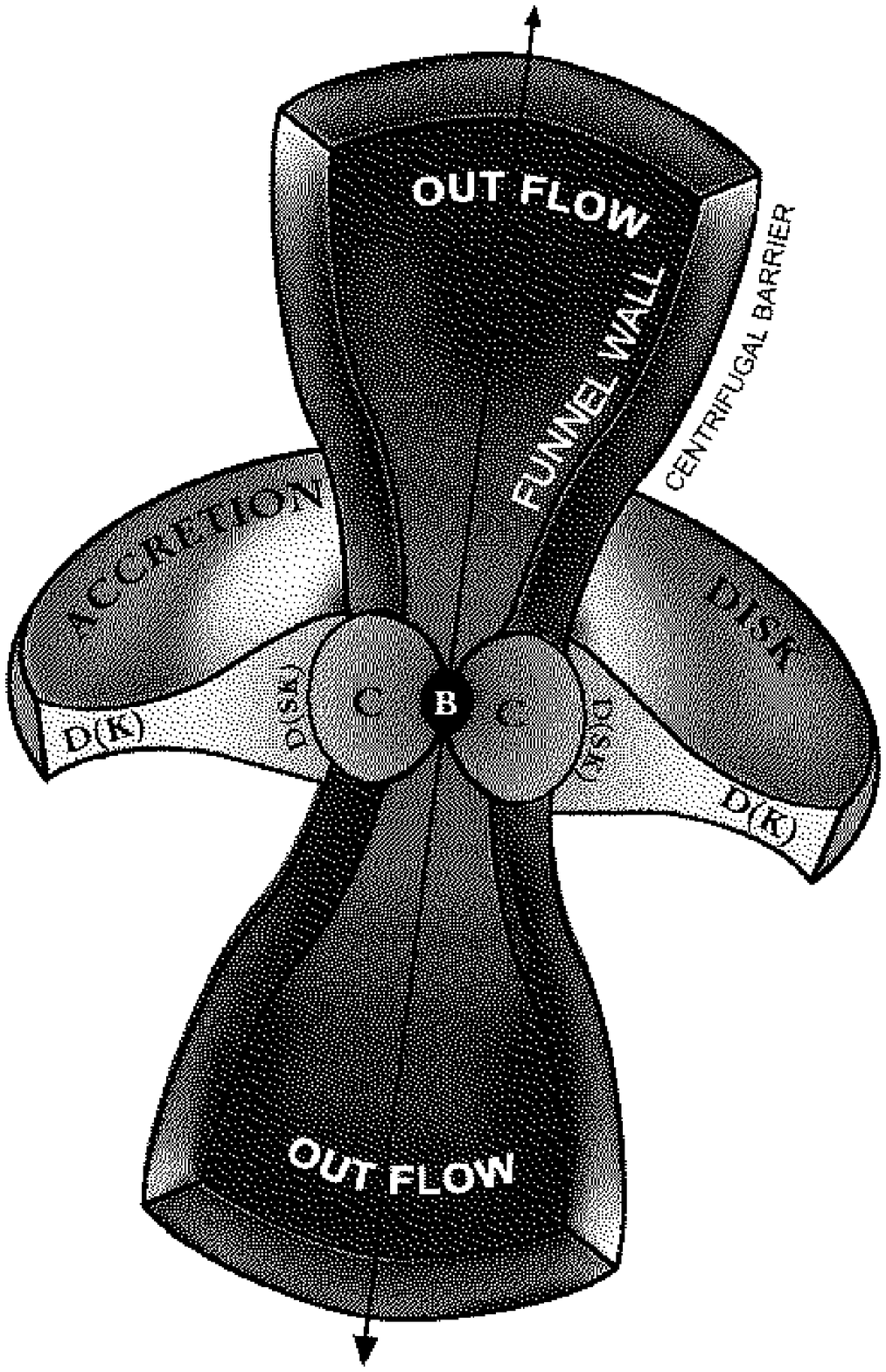}}}
\noindent {{\bf Fig. 2:}
Multi-component combined flow geometry in 3-Dimension}
\end{figure}
As we explained, the
viscous time scale is always large compared with the infall
time scale, and thus angular momentum remains almost constant and
produces the CENBOL. We are not assuming that usual Keplerian
disks (farther away) do not form. We think that while forming this Keplerian
disk, the angular momentum has been re-distributed and excess has been
absorbed by the companion as in the usual scenario. What is new
here is the realization that the angular momentum distribution
must deviate from a Keplerian distribution close to the black hole and
therefore produces a CENBOL closer to the black hole.
Such a picture of the
inflow with Keplerian flow in the high viscosity region of equatorial
plane and low viscosity flow away from it is not due to convenience
or necessity, but it arises out of the actual solutions of the viscous
transonic flow equations (see Chakrabarti \& Titarchuk 1995 and
references therein). This has not only changed our
view of the state changes of the black hole spectra (hard/soft),
this has also given us clues to the power-law tail of the soft
state etc. through the bulk motion Comptonization etc.
The interesting
feature is that the {\it same} CENBOL may be responsible for the outflow
as well. \\
\noindent
There are two surfaces of utmost importance in
flows with angular momentum. One is the
`funnel wall' where the effective potential $\Phi^i_{eff}(r)$
vanishes. In the case of a purely rotating
flow, this is the `zero pressure' surface. Flows can not enter inside the
funnel wall because the pressure would be negative.
The other surface
is called the `centrifugal barrier'. This is the surface
where the radial pressure gradient of a purely rotating flow
vanishes and is located
{\it outside} the funnel wall simply because the flow pressure is
higher than zero
on this surface. Flow with inertial pressure easily crosses this `barrier' and
either enters into a black hole or flows out as winds depending on its initial
parameters. It is observed
that (DC) the outflow generally hugs the `funnel wall' and goes
out in between these two surfaces.
Here it is important to note that
our assumption of thin inflow is for the
sake of computation of the thermodynamic quantities only, but the
flow itself need not be physically thin. Also the funnel wall
and the centrifugal barrier are purely geometric surfaces, and they exist
anyway and the outflow could be supported even by ambient medium
which may not necessarily be a part of the disk itself.
It is true that the surfaces
could have been constructed more accurately if the viscous and
other dissipative forces could also be used to obtain this. But the error
cannot be high, since the radial motion in the outflow is very
small compared to the azimuthal motion at least up to the sonic
point.\\
Qualitatively, one might attempt to `visualize' how the combined accretion-outflow
system along with the central accretor would `look like' in reality.
In figure 2., 
we attempt to illustrate the
3-D geometry of coupled disk-outflow
system according to our model. {\bf B} is the accreting Schwarzschild
black hole while {\bf C}'s represent the hot and dense CENBOL region.
{\bf D(K)} and {\bf D(SK)} represent the thin Keplerian part and puffed-up
sub-Keplerian part of the advective accretion disk respectively
(diagram not in scale). Due to the axisymmetry assumption in
accretion, two oppositely directed jet are expelled from the close vicinity
of {\bf B}. The inner and the outer surfaces of the outflow are the
funnel wall and centrifugal barrier respectively as explained above, see eq. (9a-9b) in 
next paragraph.\\
\noindent
In ordinary stellar mass loss computations,
the outflow is assumed to be
isothermal till the sonic point (Tarafdar 1988, and references therein).
This assumption is probably justified,
since copious photons from the stellar atmosphere deposit momenta on the
slowly outgoing and expanding outflow and possibly make the flow close to 
isothermal.
This need not be the case for outflows from compact sources. Centrifugal
pressure supported boundary layers close to the black hole are very hot
(close to the virial temperature) and most of the photons emitted may be
swallowed by the black holes themselves instead of coming out of the
region and depositing momentum onto the outflow. Thus, the outflows could be
cooler than isothermal flows. In our work, we choose polytropic outflows
with the same energy as the inflow (i.e., no energy dissipation between the
inflow and outflow) but with a different polytropic index $\gamma_{o} <\gamma$.
We have approximated $\gamma_o$ as a free parameter
though in reality $\gamma_{o}$ is directly related to the heating and 
cooling processes of the outflow.
However, investigation of such detailed radiative processes are beyond the
scope of this work.
When ${\dot M}_{in}$ is high, heating of outflow by photon momentum 
deposition is higher,
and therefore $\gamma_{o} \rightarrow 1$ letting the flow to approach towards
its isothermal limit.
We also assume that very little viscosity is present in the flow except at the
place where the shock forms, so that the specific angular momentum $\lambda$
is constant in both inflows and outflows close to the black hole.
That viscous time scales are longer compared to
the inflow time scale, may be a good  assumption in the disk, but it may not
be a very good assumption for the outflows which are slow prior to the 
acceleration and are therefore, prone to viscous transport of angular 
momentum. Such detailed study has not been attempted here particularly 
because we know very little about the viscous processes taking place
in the pre-jet flow. Therefore, we concentrate only those cases where the 
specific angular momentum is roughly constant when inflowing matter 
becomes part of the outflow. 
At the shock,
entropy is generated
and hence the outflow is of higher entropy for the same specific energy.\\
\noindent
For outflow, the energy conservation, mass conservation and the 
entropy conservation equations can be written as:
$$
{\cal E}=\frac{v^2}{2}+\frac{a^2}{{\gamma}_o-1}+\frac{\lambda^2}{2r^2_m}
+\Phi_i(r)
\eqno{(8a)}
$$
$$
{\dot M}={\rho}v{{\cal A}(r)}
\eqno{(8b)}
$$
$$
{\dot {\cal M}}=a^{\frac{2}{\gamma_o-1}}v{{\cal A}(r)}
\eqno{(8c)}
$$
where $v$ is the outflow velocity.\\
\noindent
The difference between  eq. (3a-3c) and (8a-8c)
is that, presently,
the rotational energy term contains
$$
{r_m}(r)=\frac{{\cal R}(r)+R(r)}{2},
$$
as the mean {\it axial} distance of the flow. The expression of ${{\cal R}(r)}$,
the local radius of the centrifugal barrier comes from balancing
the centrifugal force with the gravity:
$$
{{\cal R}(r)}=\left(\frac{\lambda^2r}{\Phi^{'}_i(r)}\right)^{\frac{1}{4}}
\eqno{(9a)}
$$
And the expression for $R(r)$, the local radius of the funnel wall,
comes from vanishing of total effective potential, i.e.:
$$
R(r)=\lambda\sqrt{\frac{\phi_i(r)}{2}}
\eqno{(9b)}
$$
so we can write the mean axial distance $r_m$ as:
$$
r_m=\frac{1}{2}\left[\lambda\sqrt{\frac{\phi_i(r)}{2}}
+\left(\frac{\lambda^2r}{\Phi^{'}_i(r)}\right)^{\frac{1}{4}}
\right]
\eqno{(9c)}
$$
where $\phi_i(r)=-\Phi_i(r)$. 
Here, ${\cal A}(r)$ is the area between the centrifugal barrier and the funnel
wall. This is computed with the assumption that the outflow is external pressure
supported, i.e., the centrifugal barrier is in pressure
balanced with the ambient medium. Matter, 
if pushed hard enough, can cross centrifugal barrier
in black hole accretion. An outward thermal force
(such as provided by the enormous post shock temperature) 
in between the funnel wall
and the centrifugal barrier causes the flow to come out. Thus the
cross section of the outflow is,
$$
{\cal A}(r)={\pi}{\lambda}\left(\frac{\lambda\phi_i(r)}{2}
+\sqrt{\frac{r}{\Phi^{'}_i(r)}}\right)
\eqno{(9d)}
$$
Like accretion, the velocity gradient of the {\it outflow} can be 
computed as:
$$
\left(\frac{dv}{dr}\right)_i=
\frac
{
\left[
\frac
{{\lambda^3{\phi^{'}_i(r)}}\sqrt{\frac{2}{\phi_i(r)}}
+\left(\frac{r\lambda^{10}}{\Phi^{'}_i(r)}\right)^{\frac{1}{4}}
\left(\frac{1}{r}-\frac{\Phi^{''}_i(r)}{\Phi^{'}_i(r)}\right)}
{
\left[\lambda\sqrt{\frac{\phi_i(r)}{2}}
+
\sqrt{\lambda}\left(\frac{r}{\Phi^{'}_i(r)}\right)^{\frac{1}{4}}\right]^3
}
\right]
+
a^2\left[
\frac
{\frac{\lambda\phi^{'}_i(r)}{2}+\frac{1}{2}\sqrt{\frac{r}{\Phi^{'}_i(r)}}
\left(\frac{1}{r}-\frac{\Phi^{''}_i(r)}{\Phi^{'}_i(r)}\right)}
{
\frac{\lambda\phi_i(r)}{2}+\sqrt{\frac{r}{\Phi^{'}_i(r)}}
}
\right]
-\Phi^{'}_i(r)
}
{\left(v-\frac{a^2}{v}\right)}
\eqno{(10a)}
$$
for which the sonic point conditions comes out to be:
$$
v^i_s=a^i_s=
\sqrt{
\left(
\frac
{
\frac{\lambda\phi_i(r)}{2}+\sqrt{\frac{r}{\Phi^{'}_i(r)}}
}
{\frac{\lambda\phi^{'}_i(r)}{2}+\frac{1}{2}\sqrt{\frac{r}{\Phi^{'}_i(r)}}
\left(\frac{1}{r}-\frac{\Phi^{''}_i(r)}{\Phi^{'}_i(r)}\right)}
\right)_s
\left(
\Phi^{'}_i(r)-
\frac
{{\lambda^3{\phi^{'}_i(r)}}\sqrt{\frac{2}{\phi_i(r)}}
+\left(\frac{r\lambda^{10}}{\Phi^{'}_i(r)}\right)^{\frac{1}{4}}
\left(\frac{1}{r}-\frac{\Phi^{''}_i(r)}{\Phi^{'}_i(r)}\right)}
{
\left[\lambda\sqrt{\frac{\phi_i(r)}{2}}
+
\sqrt{\lambda}\left(\frac{r}{\Phi^{'}_i(r)}\right)^{\frac{1}{4}}\right]^3
}
\right)_s
}
\eqno{(10(b)}
$$
One can solve the following equation to obtain the {\it outflow} sonic 
point(s):
$$
{\cal E}-\frac{1}{2}\left(\frac{\gamma_o+1}{\gamma_o-1}\right)
\left(
\frac
{
\frac{\lambda\phi_i(r)}{2}+\sqrt{\frac{r}{\Phi^{'}_i(r)}}
}
{\frac{\lambda\phi^{'}_i(r)}{2}+\frac{1}{2}\sqrt{\frac{r}{\Phi^{'}_i(r)}}
\left(\frac{1}{r}-\frac{\Phi^{''}_i(r)}{\Phi^{'}_i(r)}\right)}
\right)_s
\left(
\Phi^{'}_i(r)-
\frac
{{\lambda^3{\phi^{'}_i(r)}}\sqrt{\frac{2}{\phi_i(r)}}
+\left(\frac{r\lambda^{10}}{\Phi^{'}_i(r)}\right)^{\frac{1}{4}}
\left(\frac{1}{r}-\frac{\Phi^{''}_i(r)}{\Phi^{'}_i(r)}\right)}
{
\left[\lambda\sqrt{\frac{\phi_i(r)}{2}}
+
\sqrt{\lambda}\left(\frac{r}{\Phi^{'}_i(r)}\right)^{\frac{1}{4}}\right]^3
}
\right)_s
$$
$$
+\frac
{\lambda}
{
\left[
\sqrt{\frac{\lambda\phi_i(r)}{2}}
+
\left(\frac{r}{\Phi^{'}_i(r)}\right)^{\frac{1}{4}}
\right]^2_s
}
+\Phi^{'}_i{\Bigg{\vert}}_s
=0
\eqno{(10c)}
$$
and the outflow velocity gradient at the sonic point(s) can be obtained by 
solving the following equation:
$$
\left(\frac{2\gamma_o-1}{\gamma_o-1}\right)
\left(\frac{dv}{dr}\right)^2_s
+2a_s\left(\gamma_o-1\right)
\left[
\frac
{
\frac{\lambda\phi^{'}_i(r)}{2}+\frac{1}{2}\sqrt{\frac{r}{\Phi^{'}_i(r)}}
\left(\frac{1}{r}-\frac{\Phi^{''}_i(r)}{\Phi^{'}_i(r)}\right)
}
{\frac{\lambda\phi^{'}_i(r)}{2}+\sqrt{\frac{r}{\Phi^{'}_i(r)}}}
\right]_s\left(\frac{dv}{dr}\right)_s-
$$
$$
\frac{1}{{\pi}\lambda}\left(
\frac{a}
{\frac{\lambda\phi_i(r)}{2}+\sqrt{\frac{r}{\Phi^{'}_i(r)}}}\right)^2_s
{\Bigg{[}}
\frac{\lambda\Phi^{'}_i(r)}{2}-\frac{1}{2}
\left(r\Phi^{'}_i(r)\right)^{-\frac{1}{2}}
\left(\frac{1}{r}+\frac{1}{2}\frac{\Phi^{''}_i(r)}{\Phi^{'}_i(r)}\right)
-\frac{1}{2}\sqrt{\frac{r}{\Phi^{'}_i(r)}}
\frac{\Phi^{''}_i(r)}{\Phi^{'}_i(r)}
{\Bigg{(}}\frac{1}{2r}-\frac{1}{3}\frac{\Phi^{''}_i(r)}{\Phi^{'}_i(r)}
$$
$$
+\frac{\Phi^{'''}_i(r)}{\Phi^{''}_i(r)}{\Bigg{)}}
{\Bigg{]}}_s
-\left(\gamma_o-1\right)
\left[
\frac
{
\frac{\lambda\phi^{'}_i(r)}{2}+\frac{1}{2}\sqrt{\frac{r}{\Phi^{'}_i(r)}}
\left(\frac{1}{r}-\frac{\Phi^{''}_i(r)}{\Phi^{'}_i(r)}\right)
}
{\frac{\lambda\phi^{'}_i(r)}{2}+\sqrt{\frac{r}{\Phi^{'}_i(r)}}}
\right]^2_s
+\frac{3}{4}\lambda^2
\frac
{
\left[
{
\lambda\phi^{'}_i(r)\sqrt{\frac{2}{\phi_i(r)}}
+
\left(\frac{\lambda^2r}{\Phi^{'}_i(r)}\right)^{\frac{1}{4}}
\left(\frac{1}{r}-\frac{\Phi^{''}_i(r)}{\Phi^{'}_i(r)}\right)}
\right]^2_s}
{\left[
\lambda\sqrt{\frac{\phi_i(r)}{2}}+\left(\frac{\lambda^2r}{\Phi^{'}_i(r)}
\right)^{\frac{1}{4}}\right]^4_s
}
$$
$$
-\frac{8\lambda^2}
{\left[
\lambda\sqrt{\frac{\phi_i(r)}{2}}+\left(\frac{\lambda^2r}{\Phi^{'}_i(r)}
\right)^{\frac{1}{4}}\right]^3_s}
{\Bigg{[}}
2\lambda\phi^{'}_i(r)\sqrt{\frac{2}{\phi_i(r)}}
\left(\frac{\phi^{''}_i(r)}{\phi^{'}_i(r)}-
\frac{1}{2}\frac{\phi^{'}_i(r)}{\phi_i(r)}\right)+
\frac{1}{r}\left(\frac{\lambda^2r}{\Phi^{'}_i(r)}\right)^{\frac{1}{4}}
\left(\frac{1}{3r}-\frac{1}{4r}\frac{\phi^{''}_i(r)}{\phi^{'}_i(r)}\right)
-
$$
$$
\frac{\phi^{''}_i(r)}{\phi^{'}_i(r)}
\left(\frac{\lambda^2r}{\Phi^{'}_i(r)}\right)^{\frac{1}{4}}
\left(
\frac{\phi^{'''}_i(r)}{\phi^{''}_i(r)}-
5\frac{\phi^{''}_i(r)}{\phi^{'}_i(r)}
-\frac{1}{4r}\right)
{\Bigg{]}}_s
+\Phi^{''}_i(r){\Bigg{\vert}}_s=0
\eqno{(10d)}
$$
The mass outflow rate $R_{\dot m}$ can now be defined as:
$$
R_{\dot m}=\frac{{\dot M}_{out}}{{\dot M}_{in}}={\bf {\Psi}}\left({\cal E},
\lambda,\gamma,\gamma_o\right)
\eqno{(11)}
$$ where ${\bf {\Psi}}$ has some complicated non-linear 
non-linear 
functional form which can {\it not}
be evaluated analytically.\\
\noindent
The following procedure is adopted to obtain a complete solution of the coupled
accretion-outflow system. For any $\left[{\cal P}_i\right]\in {\bf PTR}$
(see Fig. 1),
suppose that matter first enters through the outer sonic point and passes 
through
a shock. At the shock,
part of the incoming matter, having higher entropy 
density is likely to return back as winds through
a sonic point, other than the one it just entered.
That is because, since the outflow would be heated by photons,
and thus have a smaller polytropic constant, the flow would leave the system
through an outer sonic point different from that of the incoming solution.
Thus a combination of topologies, one from the region {\bf PTR}, and the other 
corresponding to the flow with same 
$\left({\cal E}_i,\lambda_i\right)$ but now with
$\gamma=\gamma_o$, are required to obtain a 
full solution.
Thus
finding a complete self-consistent solution boils down to finding the outer 
sonic point of the
outflow and the mass flux through it.
A supply of parameters ${\cal E}$, $\lambda$, $\gamma$ and $\gamma_{o}$ 
make
a self-consistent computation of $R_{\dot m}$ possible.
We obtain the inflow sonic point by solving Eq. (4c).
From Eq. (41-4d), using the fourth order
Runge Kutta method, $u(r)$, $a(r)$ and
the inflow Mach number
$\left[\frac{u(r)}{a(r)}\right]$ are computed along with the  inflow from the
{\it inflow} sonic point
till
the position where the shock forms. The shock location is calculated
by solving Eq. (5). Various shock parameters
(i.e., density, pressure etc at the shock surface) are then computed
self-consistently. For outflow, with the known value of ${\cal E}$,$\lambda$ 
and $\gamma_o$, it is easy to compute the location of the outflow sonic point
from Eq. (10c). At the
outflow sonic point, the outflow velocity $v^i_s$ and polytropic sound
velocity $a^i_s$ is computed
from Eq. (10b). Using Eq. (10a) and (10d),
$\left(\frac{dv}{dr}\right)$
and $\left(\frac{dv}{dr}\right)_s$ is computed as was
done for the inflow.
Runge
-Kutta method is then
employed to integrate from the {\it outflow} sonic point
towards the black hole
to
find out the outflow velocity and density at the shock location for a given
accretion rate and ${\dot M}_{out}$ is 
calculated from Eq. (8b). The mass outflow rate $R_{\dot M}$ is then computed 
using Eq.(11) and the corresponding $\nu_{QPO}$ is computed using
Eq. (7).\\
\noindent
As $\Phi_1$ is the best available potential to mimic the general
relativistic results for multi-transonic shocked accretion, we will 
calculate $\nu_{QPO}$ and  $R_{\dot M}$ only for flows in $\Phi_1$
by setting:
$$
\left[\Phi_i(r),\Phi^{'}_i(r),\Phi^{''}_i(r),\Phi^{'''}_i(r),\right]
~{\Longrightarrow}~
\left[\Phi_1(r),\Phi^{'}_1(r),\Phi^{''}_1(r),\Phi^{'''}_1(r),\right]
$$
while simultaneously solving equations (2-11), using the procedure described 
above. One may ask the question that if we plan to explore the 
$\nu_{QPO}~-~R_{\dot M}$ correlation only for $\Phi_1(r)$, why then we formulate 
all the equations for all $\Phi_i(r)$s in general instead of 
$\Phi_1(r)$, which would be a relatively less complicated procedure. Our 
answer to this question is mainly two-fold. Firstly, we wanted to present a
generalized formulation for {\it all} $\Phi_i(r)$s to get convinced that 
\begin{figure}
\vbox{
\vskip 0.0cm
\hskip 0.0cm
\centerline{
\psfig{file=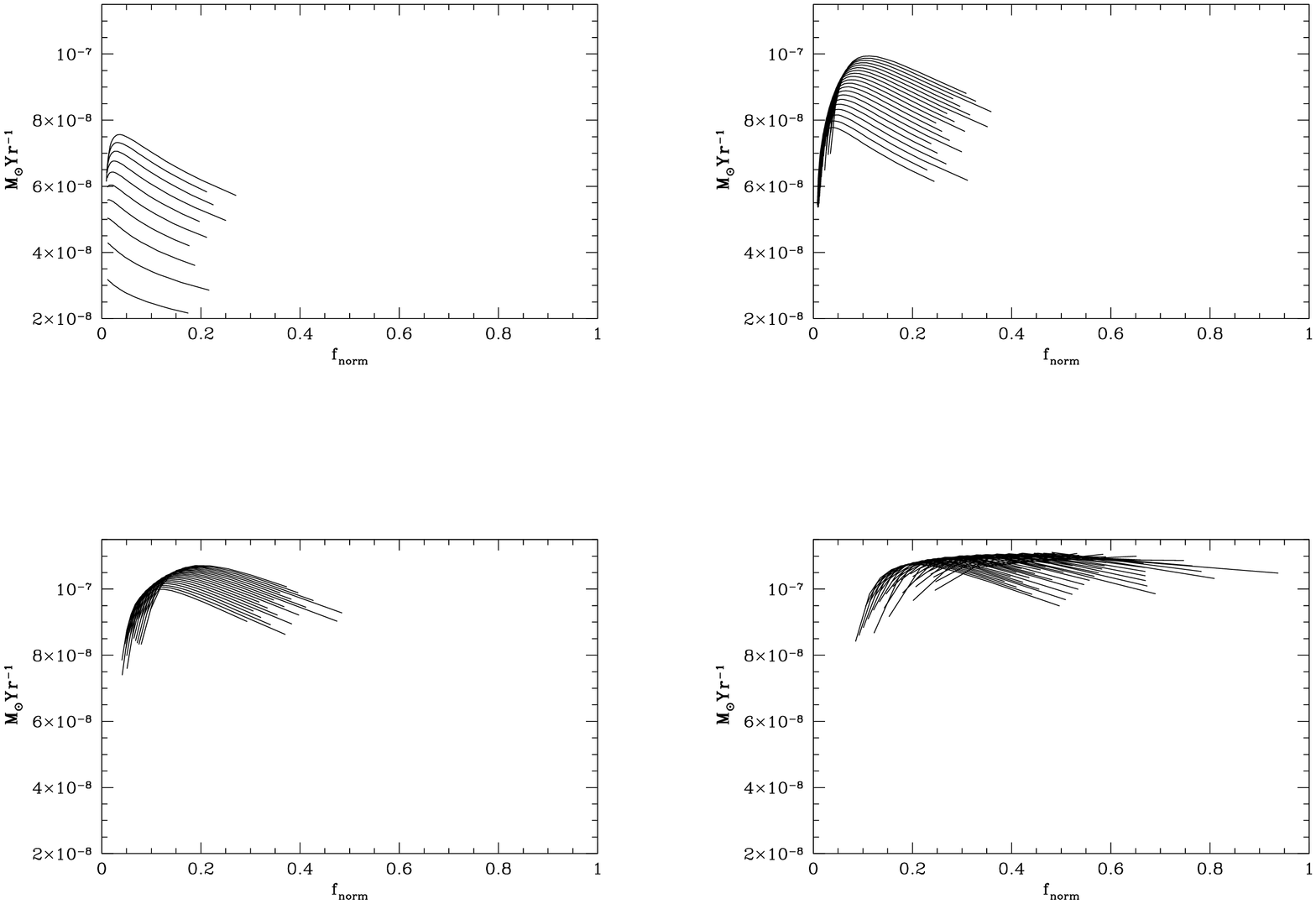,height=18cm,width=20cm,angle=270.0}}}
\noindent
{{\bf Fig. 3}
Correlation among the QPO frequencies and the mass outflow rates
for coupled accretion-wind system in $\Phi_1(r)$. ${\bf f_{norm}}=
\frac{\nu}{\nu_{max}}$ is the normalized dimensionless QPO frequency
which is basically the 
ratio of the theoretically calculated QPO frequencies to the 
highest possible observed QPO frequency for any particular 
astrophysical source, see text for detail.}
\end{figure}
\begin{figure}
\vbox{
\vskip 1.5cm
\centerline{
\psfig{file=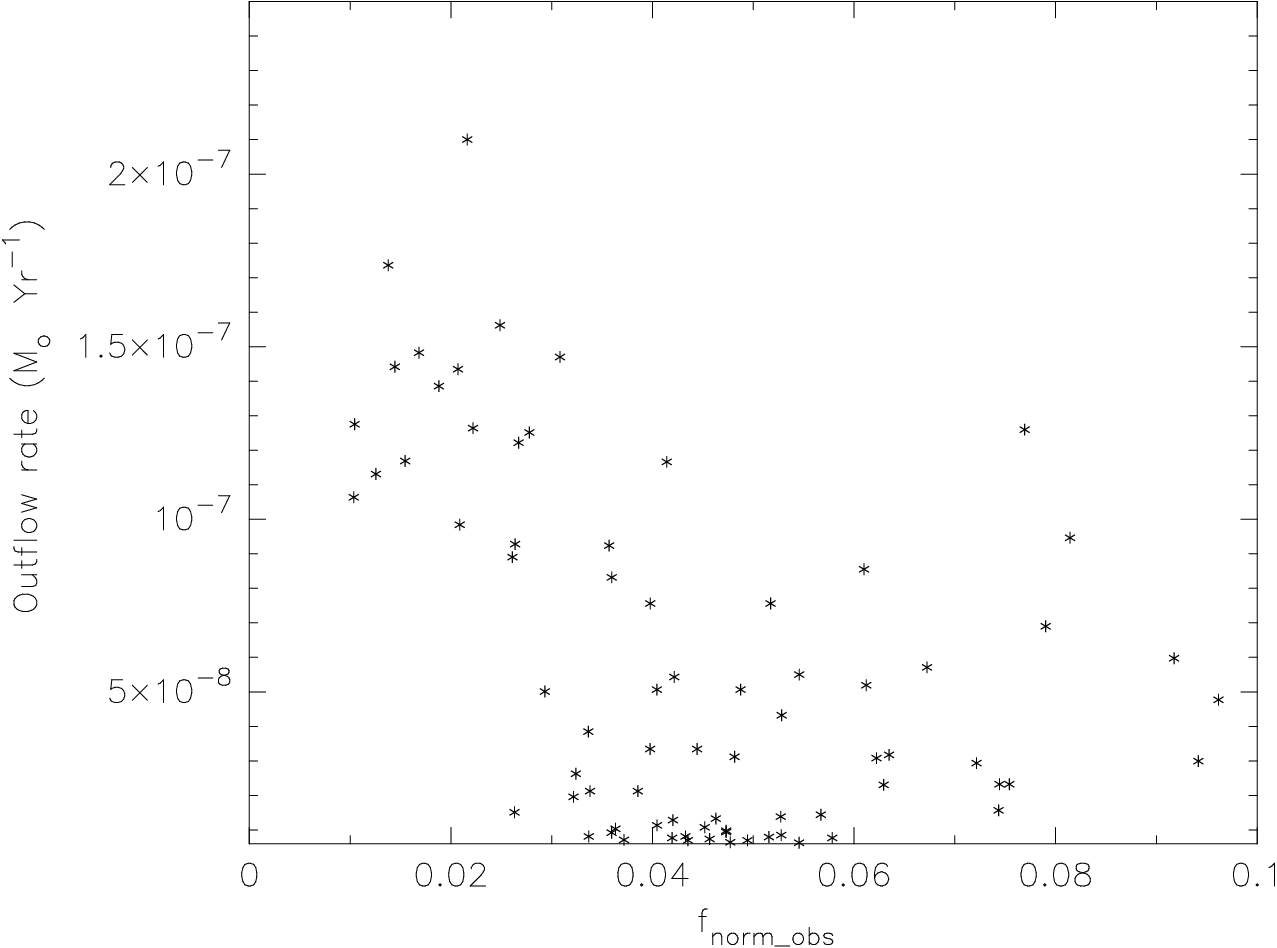,height=12cm,width=12cm,angle=-90.0}}}
\noindent {{\bf Fig. 4:}
The outflow rate plotted against normalized QPO frequency for 
GRS 1915+105, as calculated from the observed radio flux and
the measured QPO frequency.
}
\end{figure}
such correlation can be theoretically investigated for {\it any}
kind of black hole potential and to assure that our model is 
{\it not} just an artifact of a particular type of potential only.
Secondly, of course there are possibilities that in future someone 
may come up with a pseudo-Schwarzschild potential better than $\Phi_1(r)$,
which will be the best approximation for complete general relativistic 
investigation of multi-transonic shocked flow. In such case, if one can
in advance formulate a generalized model for coupled accretion-wind
system for any arbitrary $\Phi_i(r)$, exactly what we have done in this paper,
then that generalized model will be able to readily accommodate that new 
$\Phi(r)$ without having any significant change in the fundamental 
structure of the formulation and solution scheme of the model
and we need not
have to worry about providing any new scheme exclusively valid only for
that new potential, if any. Thus we believe that our general inflow-outflow
model presented in this paper, which bears the same philosophy as provided by
D98 and DC but treats the problem in a much more general way, will be
quite useful for any further improvement in this field and our theoretical
correlation study can also be extended easily for any such new and more accurate
potential that might be discovered in future.
It is to be noted that we cannot accelerate any matter more than the sound
speed (or, the rotational velocity) at the initial injection point so we
do not claim to explain superluminal motions.
We have done what is hydrodynamically possible. If we follow our
path for MHD flows, we believe that we could reach superluminal motions
also. 
Also, we
could not do a thorough job on angular momentum distribution {\it inside}
the jet. What we understand here is that in case, some form of viscosity
is working (say radiative viscosity) inside the jet to make the
distribution of the angular momentum as a power-law, then the
resulting density will go up towards the edge of the jet (a `hollow'
jet as in Hawley, Smarr \& Wilson 1984, for example).
Thus, since maximum amount of matter is taking away maximum amount of
angular momentum (unlike in the disc), the average angular momentum
need not be different from that of the CENBOL region. But in case,
the average angular momentum were lower, then we find that the
outflow rate was also lower. \\
\noindent
In Fig. 3, we represent our theoretical results
regarding the correlation of ${\nu}_{QPO}$ with $R_{\dot m}$. We define the
normalized QPO frequency $f_{norm}$ in the following way:
$$
f_{norm}=\frac{\nu}{{\nu}_{max}}
\eqno{(12)}
$$
where ${\nu}={\nu}_{QPO}$ is the theoretically calculated QPO
frequency obtained from eqn. (7), with properly scaled
$C$ 
and  ${{\nu}_{max}}$ is the
highest possible observed frequency for any particular
astrophysical source.
This
normalization is necessary because exact value of $C$ in the equation
cannot be estimated analytically. We plot $f_{norm}$ along the $X$
axis and the mass outflow rate scaled in solar mass
per year is plotted along the $Y$ axis. We calculate the
${\nu}_{QPO}$ as well $R_{\dot m}$ for such shocked ultra-relativistic
accretion on to a 10 $M_{\odot}$ black hole 
at unit Eddington rate ${\dot M}_{Edd}$.
The value of ${\cal E}$ and $\lambda$ for which the calculation is
performed are taken from the region {\bf PTR} of Fig. 1. The
scheme for the calculation is the following: \\
\noindent
First we calculate shock location $r_{sh}$, shock
compression ratio $R_{comp}$ and consequently the QPO frequency
${\nu}_{QPO}$ for the fixed lowest possible ${\cal E}$ with the range of
$\lambda$ for which the shock forms. We also calculate $R_{\dot m}$ for
this set of parameters. We then repeat this
calculation for higher values of energy
till all ${\cal E}$ and $\lambda$ covering the region {\bf PTR} are
exhausted.\\
\noindent
Fig. 3 is divided into four blocks. Each
individual curve in any block represents the variation
of $f_{norm}$ on $R_{\dot m}$ resulted from the variation of $\lambda$ by
keeping ${\cal E}$ fixed, and different curves are drawn
for different ${\cal E}$ with ${\Delta}{\cal E}=10^{-3}$. 
As stated earlier, ${\cal E}$ and $\lambda$
covering the region {\bf PTR} are used to draw Fig. 3.
The whole range of specific energy of the shocked
accretion material for which computation is done, was
divided into four parts. The top left block represents
results from very low energy accretion, top right for medium energy
accretion, bottom left for moderately high energy
accretion and bottom right is for very high energy
accretion. One can observe that for very low energy
accretion, normalized QPO frequency  non-linearly and
monotonically anti-correlates with the mass outflow
rate and the calculated ${\nu}_{QPO}$ is also relatively low. As
the specific energy of the accreting material
increases, a peak appears in the correlation curve
and the monotonic behavior is not maintained anymore.
As  ${\cal E}$ becomes higher, the normalized QPO
frequency first nonlinearly correlates with the mass
outflow rate up to a certain value of ${\nu}_{QPO}$ (or with $f_{norm}$)
for
which the mass loss rate becomes maximum, then the
anti-correlation phase starts like low energy accretion cases.\\
\noindent
It is to be noted here that both $f_{norm}$ as well as $R_{\dot m}$ depend on
initial boundary conditions and are quite sensitive to the shock location $r_{sh}$.
while $f_{norm}$ non-linearly and monotonically anti-correlates with $r_{sh}$, one
can show that $R_{\dot m}$ represents a distinctive non-monotonic variation with
$r_{sh}$. When shock location is small, $R_{\dot m}$ nonlinearly correlates with
$r_{sh}$ and attains a peak a particular value of $r_{sh}$, say at $r_{sh}^{p}$. For
$r_{sh}~>~ r_{sh}^p$ (if available for a set of initial boundary condition),
$R_{\dot m}$ starts anti-correlating with $r_{sh}$. These informations might be used
in the following way to infer whether our calculation is model dependent or not.\\
For the same set of initial boundary conditions, the shock location distinctively
differs for various $\Phi_i$ used here. This is because of the fact that the region
of parameter space (spanned by ${\cal E}$, $\lambda$ and $\gamma$) responsible for
shock formation is quite different for almost all four $\Phi_i$ (see Fig. 4 and Fig.
7 of D02). Hence for a particular value of $[{\cal P}_i]$, one obtains {\it numerically
different} value of $f_{norm}$ and $R_{\dot m}$ for all $\Phi_i$ in general.
This information, by any means, does {\it not} imply that our calculation is model
dependent \footnote {By the term `model (in) dependent; we mean whether the basic
physics remains unaltered or drastically changes with a specific choice of $\Phi_i$.}
because though the numerical values may differ, the overall $f_{norm}-R{\dot m}$
(anti) correlation profile {\it does} remain the same for all $\Phi_i$. This is because of the fact
that $f_{norm}-r_{sh}$ anti-correlation and $R-{\dot m}-r_{sh}$  (anti) correlation profile
remains unaltered for all
$\Phi_i{s}$, only the numerical values of the related quantities change with the choice
of potential. Thus we may, perhaps, conclude that our calculation is more or less
`generalized' in nature to include all existing pseudo-Schwarzschild potentials.

\section{Comparison with observations}

 The Galactic microquasar GRS 1915+105 shows several properties which 
 are indicative of the existence of shock oscillations in terms of QPOs
 and collimated jets in terms of radio emission (see Belloni 2002 for
 a general review on this particular source). 
 Since it has showed intense and highly variable X-ray and radio emission 
 characteristics, it has been extensively studied and it is an
 ideal source to test our concepts on shock oscillations and
 outflows.

 GRS 1915+105 shows diverse variability characteristics (see Belloni et al.
 2000a for a classification) and the inter-relationship between these
 variability classes and the radio emission has been examined in detail
 (Naik \& Rao 2000; Belloni et al. 2000b; Muno et al. 2001; Rao et al.
 2000). Highly variable steep spectrum radio emission are invariably
 associated with superluminally moving ejecta and these are causally
 connected to a
 disturbed accretion disc as observed in X-rays as some sort
 of ``inner disk oscillations''. These events could be initiated by
 some catastrophic events like the ``magnetic rubber band'' effect
 (Nandi et al. 2001). Steady flat spectrum radio emission, on the other
 hand, is  always associated with the hard X-ray state of GRS 1915+105
 and one of the identifying feature of such hard states is the
 existence of an ubiquitous 0.5 -- 10 Hz QPO (Muno et al. 2001). 
 It would be quite instructive to compare our results obtained
 in the previous sections to the outflow properties (as observed
 in radio) and shock oscillation properties (measured as the
 0.5 -- 10 Hz QPO) of GRS 1915+105.

  When the outflow is confined to the outer sonic point in the
  wind, the material can get collected and after some time can cause
  a catastrophic Compton cooling. This concept has been successfully
  used by CM to explain the repeated 
  variation between high and low states in GRS 1915+105 and
  a relation between the QPO frequency and the `low-state' (or
  `off' state) duration has been derived and compared with the 
  observations. Based on this concept, we can relate the 
  predicted QPO frequency to the observed one by scaling to the
  maximum observed QPO frequency of 67 Hz (Morgan et al. 1997).

 To convert the observed radio luminosity to outflow rate, we adapt the
 method given in Fender \& Pooley (2000). We assume that
 the radio spectrum has a flat spectrum from 1 GHz to 1.4$\times$10$^5$
 Hz (corresponding to mid-IR region of 2.3$\mu$m) and get a conversion
 factor of observed radio flux to jet luminosity as 1.35$\times$10$^{35}$
 erg s$^{-1}$ per mJy (for a distance of 12.5 kpc). The jet 
 luminosity, $L_J$, can be converted to jet power, $P_J$ by 
 $$
 P_J \sim L_J \eta^{-1} F(\Gamma,i)
 \eqno{(13)}
 $$
where $\eta$ is the radiative efficiency and $F(\Gamma,i)$ is the correction
factor for bulk motion (see Fender \& Pooley 2000). $\eta^{-1} F(\Gamma,i)$ has
a typical value of 10. Further assuming that the jet emitting region 
has a typical volume of 10$^{40}$ cm$^3$ and a magnetic field of
100 G, we can convert the jet power to outflow rate, getting a
conversion factor of 1.4 $\epsilon$ $\times$ 10$^{-8}$ M$_{\odot}$ yr$^{-1}$ per 1 mJy
for GRS 1915+105, where $\epsilon$ is the fraction of particles accelerated to
relativistic energies. In our present work, we have calculated the total outflow
and further acceleration, presumably in some shock regions in the
jet, would be required to accelerate these particles to relativistic
energies. Since our purpose in the present work is only to get
an estimate of the relative variation of outflow rate to shock oscillations,
we have taken a conservative value of 0.1 for $\epsilon$.

Using these conversion factors, we have plotted the observed outflow rate
 as a function of normalized QPO frequency, f$_{norm-obs}$ in Figure 4, based
 on a compilation by Muno et al. (2001).
 It should be noted that there is a general agreement with the
 trend of observations as predicted by the theory (see Figure 3)
 in the sense that the outflow rate shows an increase with
 f$_{norm}$ at low frequencies,
 reaches a maximum, and then slowly declines.
 The agreement, however, is
 particularly close for the case of low specific energy (Figure 3, top left).
 The QPOs seen in GRS 1915+105 pertains to the steady low-hard states
 where the accreted matter can steadily loose its energy till it
 reaches the CENBOL and the specific energy can be very low
 for the accreting matter close to the black hole. The system can go into
 a high specific energy accretion mode at certain occasions (for example
 during the infrequent times when the
 high frequency - 67 Hz -  QPOs are seen).
 It would be interesting to steady
 the behavior of QPO frequency and the outflow rate during such modes
 and we speculate that the variations would
 look more like the high specific energy behavior (Figure 3, bottom
 right).  Also, there are possibilities that the outflowing matter (the 
ejecta) is essentially dominated by low binding energy (because they
are supposed to be escaped towards infinity), hence such co-rrelation 
in stronger for low energy accretion.
 We must, however, caution that there are several uncertainties in the
 derivation of the outflow rate from the observed radio luminosity,
 but, the connection between the radio emission and the properties of
 the accretion disc seems to be on the line predicted by the
 Advective Flow formalism, described in detail in this paper. A detailed
 simultaneous observations and a calculation of the X-ray spectra using
 proper hydrodynamic formalisms
 would be required to further refine the model predictions.

It is worth pointing out that the total amount of outflow observed
in GRS~1915+105 during the `baby-jet' episodes as well as during
the super-luminal jet emission are also consistent with calculations
presented here. Multi-wavelength observations of GRS~1915+105 (Mirabel
et al. 1998) showed evidence for a blob of matter being ejected out,
which are called the `baby-jets', and it is estimated that $>~10^{19}$ g
of matter is
ejected during these events. It was pointed out by Belloni et al.
(2000a) that these events are always preceded by hard dips characterized
by $\sim$3 Hz QPOs and they hypothesis that the matter for jet emission
is ejected during these hard dips lasting for $\sim$500 s. Hence,
a mass outflow of $\sim$10$^{-10}$ M$_\odot$ yr$^{-1}$ must be
participating in the relativistically accelerated particles, which is
about 1\% of the outflow estimated in our calculations.
GRS~1915+105 also shows superluminal jet emission with an estimated total
mass of 10$^{23}$ g (Mirabel \& Rodriguez 1999; Fender et al. 1999).
These
events could be either due to a series of dips preceded by low hard
states
with QPO frequency $\sim$ 1 Hz (Naik et al. 2001) or due to the
accumulation
of matter during an enhanced low state lasting for a few days (Dhawan et
al. 2000).
In either case, there is evidence that the ejected matter is accumulated
during the low hard state with equivalent f$_{norm}$ of 0.01 and lasting
for $\sim$50,000 s giving an effective outflow rate of 10$^{-8}$ 
$M_{\odot}~yr^{-1}$,
consistent with the calculation presented here.
\section{Concluding Remarks}
In this paper we attempt to calculate the QPO frequency of the central
accretor harbored by the galactic microquasar sources and to explore the 
dependence of the amount of baryonic load in the microquasar jets
on these frequencies. Our calculation has been performed for 
{\it polytropic} shocked accretion onto galactic stellar
mass black holes using a generalized post-Newtonian pseudo-Schwarzschild
framework. However, one can show that shock 
formation is also possible when accretion is taken to 
be {\it isothermal} (Das, Pendharkar \& Mitra 2003), and it is
possible to calculate the normalized QPO frequencies for such accretion 
flow as well (Das 2003). In our future work, we would like to explore 
the correlation between $\nu_{QPO}$ and $R_{\dot m}$ for shocked isothermal 
inflow as mentioned above. \\
Theory of accretion onto black holes is inherently a tough and
intractable problem: first of all the nature of viscosity
which is known to be turbulent even in mildly relativistic
accretion flows is difficult to even formulate, let alone solve,
in the relativistic accretion flows likely to be present in
these system and the inner boundary condition is impossible to
formulate and analytically handle in regions where general theory of
relativity is
known to predominate. The approach that  we have taken in this work
is the
one  in which general theory of relativity
is handled in a set of pseudo-Schwarzschild
potentials and the solutions are sought in regimes close to
the black hole where the flow is shown to be supersonic and
likely to be inviscid.
Two fundamental problems are inherent in this approach namely:\\
\noindent
1) How are we sure that the pseudo-Schwarzschild potentials
   give correct description of general theory of 
relativity,\\
\noindent
and\\
\noindent
2) How do we extrapolate the
   results from inner regions of accretion disc where the accretion
   is likely to be inviscid to the outer regions where the flow is
   viscous and most of the observables like X-ray emission characteristics
   are measurable. \\
\noindent
We understand that such questions cannot be
   answered with mathematical uncertainty, but the present paper
   takes very important steps in that direction.
By showing that
   shocks exist in all available pseudo-Schwarzschild potentials
   we have increased the confidence that shocks should indeed exist
   in the correct general relativistic 
formalism. By carefully selecting some
   important observables which are likely to be the manifestations
   of the conditions of inner accretion disk, we have attempted
   to make a semi-qualitative comparison with the theoretical
   predictions and have obtained very encouraging results. Though the
   fundamental problem of obtaining a complete solution for
   accretion onto black holes is not solved, very important insights 
   are obtained 
   in the present work which increases the confidence
   in the methodology and detailed  prescriptions are given  for comparison
   with observations.
\section{Acknowledgements}
This research has made use of NASA's Astrophysics Data System Bibliographic Services.
Research of TKD at UCLA is supported by NSF funded post doctoral
fellowship (Grant No. NSF AST-0098670).
TKD would like to acknowledge the hospitality provided by 
the Racah Institute of Physics, The Hebrew University of Jerusalem, 
Israel, where a part of this work had been done. He would also like to
acknowledge the hospitality provided by the
Department of Astronomy and Astrophysics,
Tata Institute of Fundamental Research, where he used to visit frequently to
work with the members of the X-Ray Astronomy Group.
The authors would like to thank the annonimeous refree for a number of
useful comments.

\end{document}